\documentclass[twocolumn,showpacs,amsmath,amssymb,aps,prb]{revtex4-1}
\usepackage{graphicx}
\usepackage{dcolumn}
\usepackage{bm}
\usepackage[abs]{overpic}
\usepackage{color}

\def\up{\uparrow}
\def\down{\downarrow }

\begin{document}

\title{Self-consistent study of Abelian and non-Abelian order in a two-dimensional topological superconductor}

\author{S. L. Goertzen}
\affiliation{Department of Physics and Engineering Physics, University of Saskatchewan, 116 Science Place, Saskatoon, SK, S7N 5E2 Canada}

\author{K. Tanaka}
\email{Corresponding author: kat221@campus.usask.ca}
\affiliation{Department of Physics and Engineering Physics, University of Saskatchewan, 116 Science Place, Saskatoon, SK, S7N 5E2 Canada}

\author{Yuki Nagai}
\affiliation{CCSE, Japan  Atomic Energy Agency, 178-4-4, Wakashiba, Kashiwa, Chiba, 277-0871, Japan}

\date{\today}

\begin{abstract}
We perform self-consistent studies of two-dimensional (2D) $s$-wave topological superconductivity (TSC) with Rashba spin-orbit coupling and Zeeman field by solving the Bogoliubov-de Gennes equations. In particular, we examine the effects of a nonmagnetic impurity in detail and show that the nature of the spin-polarised midgap bound state varies significantly depending on the material parameters. Most notably, a nonmagnetic impurity in a 2D $s$-wave topological superconductor can act like a magnetic impurity in a conventional $s$-wave superconductor, leading to phase transitions of the ground state as the impurity potential is varied. Furthermore, by solving for the spin-dependent Hartree potential self-consistently along with the superconducting order parameter, we demonstrate that topological charge density waves can coexist with TSC at half filling just as in a conventional $s$-wave superconductor.
\end{abstract}

\pacs{74.20.-z, 74.20.Rp, 74.81.-g, 75.70.Tj}
\maketitle

\section{\label{sec:introduction}Introduction}

The Rashba spin-orbit (SO) interaction\cite{Rashba1960} that occurs in materials without inversion symmetry is a key ingredient for topological insulating or superconducting states and for creating Majorana fermions as elementary excitation in a wide variety of systems.\cite{Manchon2015} Two-dimensional (2D) $s$-wave topological superconductivity (TSC) with Rashba SO coupling and Zeeman field has been proposed to be realised in an ultracold Fermi gas, where $s$-wave superfluidity and SO coupling can be generated, respectively, via the $s$-wave Feshbach resonance and spatially varying laser fields.\cite{Sato2009,Sato2010} 2D $s$-wave TSC has also been proposed to be achievable in a solid device made of more conventional materials,\cite{Sato2010,Alicea2012} such as a semiconductor with SO coupling sandwiched between a conventional $s$-wave superconductor and a ferromagnetic insulator,\cite{Sau2010} or a semiconductor with Rashba and Dresselhaus coupling in proximity to an $s$-wave superconductor only.\cite{Alicea2010} Furthermore, recently developed, one-atom-layer Tl-Pb compounds on Si(111) with giant Rashba effects\cite{Matetskiy2015} and ionic-liquid based double-layer transistors\cite{Li2016} are promising candidate systems for realising 2D $s$-wave TSC.\cite{Nagai2016} In the latter system, for example, the layered structure can consist of an $s$-wave superconductor and a ferromagnetic insulator, while the electric field controls the charge-carrier density as well as the Rashba SO interaction.\cite{Nagai2016}  

In a 2D $s$-wave topological superconductor, vortices host Majorana fermions as zero-energy bound states\cite{Sau2010,Tewari2010} and hence obey non-Abelian exchange statistics. Therefore, creating 2D $s$-wave TSC in a condensed matter system, where one has control over key parameters such as the filling factor and the strength of Rashba SO coupling and/or Zeeman field, potentially leads to realisation of fault-tolerant topological quantum computation.\cite{Alicea2012,Nayak2008} For setups where superconductivity is induced by the proximity effect, in general there have been several arguments\cite{Stanescu2010,Potter2011,Alicea2012} that coupling to the superconductor can renormalise the original parameters, e.g., on the surface of a three-dimensional topological insulator and can even be detrimental to the other ingredients necessary for TSC; although a self-consistent study of the proximity effect\cite{Black2011} has contradicted some results in this regard in Ref.~\onlinecite{Stanescu2010}. 
It is ideal to have an intrinsic pairing interaction in the system that drives $s$-wave superconductivity or superfluidity, as in $s$-wave superfluids of fermionic atoms\cite{Sato2009}, one-atom-layer superconducting compounds,\cite{Matetskiy2015} or electric-field double-layer transistors.\cite{Li2016,Nagai2016}

Despite the rapid advancement in device fabrication techniques and the 2D $s$-wave TSC model\cite{Sato2009,Sato2010,Sau2010,Alicea2012,Alicea2010} being one of the most promising models for platforms for topological quantum computation, self-consistent studies of this model are severely lacking. Few studies made so far, in which the superconducting order parameter is solved self-consistently, are a momentum-space study of average impurity effects\cite{Nagai2014oval} and studies of a single vortex\cite{Bjornson2013,Bjornson2015} and the vortex lattice\cite{Smith2016} in terms of the tight-binding model;\cite{Sato2009,Sato2010} and a study of the effects of a nonmagnetic impurity in an $s$-wave superfluid\cite{Hu2013} using the continuum model.\cite{Sau2010} To the best of our knowledge, there has been no work where the Hartree potential is solved self-consistently as well.

The purpose of the present work is to perform self-consistent studies of 2D $s$-wave TSC by solving the Bogoliubov-de Gennes (BdG) equations\cite{deGennes} on the tight-binding model of Sato, Takahashi, and Fujimoto\cite{Sato2009,Sato2010} directly and numerically. As a tight-binding model, it is versatile in that band structure and the filling factor can easily be adjusted to model real systems in terms of hopping amplitudes and the chemical potential. Moreover, unlike the continuum model,\cite{Sau2010} in addition to non-Abelian phase this model can host Abelian phase where topological order can be realised not only in superconducting states, but also in density wave states.\cite{Sato2010} In this work, we examine the effects of a nonmagnetic impurity and illustrate how the properties of the midgap bound state vary with the material parameters -- in stark contrast to the claim made by Hu {\it et al}.\cite{Hu2013} on ``universal'' midgap bound states. In particular, we find that the midgap excitation bound to a nonmagnetic impurity is spin-polarised\cite{Nagai2015} and its spin can flip as the impurity potential is varied. Most notably, we find that a nonmagnetic impurity in a 2D $s$-wave topological superconductor can act exactly like a magnetic impurity (classical spin) in a conventional $s$-wave superconductor, resulting in phase transitions of the ground state. We also demonstrate that the spin-dependent Hartree potential effectively reduces the Zeeman field. Furthermore, by solving for the Hartree potential as well as the superconducting order parameter self-consistently, we show the existence of topological charge density waves (TCDW) as the ground state that is degenerate with TSC at half filling, just as in a conventional $s$-wave superconductor. 

Sau and Demler\cite{Sau2013} have studied the effects of a nonmagnetic impurity in a semiconductor nanowire with Rashba SO coupling and Zeeman field in proximity to an $s$-wave superconductor,\cite{Lutchyn2010,Oreg2010} where one-dimensional (1D) TSC as in Kitaev's model can be realised.\cite{Kitaev2001,Alicea2012}  By searching for poles in the Green function, they found no midgap excitation bound to a nonmagnetic impurity when either one of the Rashba SO interaction and the Zeeman field -- the two ingredients required for TSC -- was missing. In TSC states, when the Zeeman field is larger than the assumed uniform order parameter, they have found midgap bound states whose energy can cross zero as the strength of the impurity potential is increased. Our self-consistent results for zero-energy crossing of the midgap excitation in a 2D $s$-wave topological superconductor are similar to their findings; except that in the 1D system, the impurity cuts the wire in half and creates Majorana edge modes in the limit of infinitely strong potential.

We employ the Chebyshev polynomial expansion method\cite{Covaci2010,Nagai2012} for solving the BdG equations self-consistently for the mean fields as well as calculating the local density of states (LDOS) after self-consistency has been achieved. This method allows one to obtain self-consistent mean fields without diagonalization of the BdG Hamiltonian matrix and it can also gain significant speed-up from parallel computation. It is thus much more efficient than the traditional way of solving the BdG equations by direct diagonalization, especially when the BdG matrix is spin-dependent and required to be solved for relatively large system size. We also circumvent the high numerical demand of diagonalizing the BdG matrix by utilising the efficient algorithm of Sakurai and Sugiura (SS) \cite{Sakurai2003,Nagai2013} to obtain the quasiparticle spectrum within a selected energy window. This numerical technique also benefits greatly from parallelism and if desired, the entire spectrum can be obtained readily by dividing the energy range into smaller subranges and applying the SS method to each subrange.

The paper is organised as follows. The model is described in Sec.~\ref{sec:model}, results are presented and discussed in Sec.~\ref{sec:results}, and the work is summarised in Sec.~\ref{sec:conclusions}.

\section{\label{sec:model}Model}

We solve the BdG equations on the tight-binding model for 2D $s$-wave TSC\cite{Sato2009,Sato2010} with the Hamiltonian,
\begin{eqnarray} 
\mathcal{H} &=& \sum_{\langle ij \rangle \sigma} t_{ij} c^\dag_{i \sigma} c_{j \sigma} + \sum_{i\sigma} (\epsilon_i -\mu + h_\sigma + V_{i \bar{\sigma}}^{(H)}) c^\dag_{i \sigma} c_{i \sigma}\nonumber\\
&+& \frac{\alpha}{2} \biggl[\, \sum_i (c^\dag_{i-\hat{x} \downarrow} c_{i \uparrow} - c^\dag_{i+\hat{x} \downarrow} c_{i \uparrow}) \nonumber \\
&+& i(c^\dag_{i-\hat{y} \downarrow} c_{i \uparrow} - c^\dag_{i+\hat{y} \downarrow} c_{i \uparrow}) + \text{H.c.}\, \biggr] \nonumber \\
&+& \sum_i (\Delta_{i} c^\dag_{i \uparrow} c^\dag_{i \downarrow} + \text{H.c.})\,,
\label{hamiltonian}
\end{eqnarray}
with the Zeeman energy $h_\sigma=-h$ and $+h$ for $\sigma=\uparrow$ and $\sigma=\downarrow$, respectively, and $\bar{\sigma}\ne \sigma$. We assume a uniform pairing interaction $U_i\equiv U$ that results in the $s$-wave order parameter $\Delta_i$ and the Hartree potential $V_{i \sigma}^{(H)}$:
\begin{eqnarray}
\Delta_{i} &=& U \langle c_{i \downarrow} c_{i \uparrow} \rangle ,\label{order}\\
V_{i \sigma}^{(H)} &=& U \langle c_{i \sigma}^\dagger c_{i \sigma} \rangle ,\label{hartree}
\end{eqnarray}
where the electron creation and annihilation operators at site $i$ with spin $\sigma$ are denoted as $c^\dag_{i\sigma}$ and $c_{i\sigma}$, respectively, and $V_{i \sigma}^{(H)}$ is the Hartree potential created by the electrons with spin $\sigma$ (and felt by those with the opposite spin $\bar{\sigma}$) at site $i$. In the Hamiltonian (\ref{hamiltonian}), we consider hopping among nearest-neighbour lattice sites $\langle ij \rangle$ only with the probability amplitude $t_{ij}\equiv -t$, $\mu$ is the chemical potential, $\epsilon_i$ is the single-particle potential due to a nonmagnetic impurity at site $i$, $\alpha > 0$ is the Rashba spin-orbit coupling strength, and H.c. stands for the Hermitian conjugate. We set the lattice constant to be unity, and $\hat{x}$ and $\hat{y}$ are the unit vectors in the $x$- and $y$-directions. 

Defining the average Hartree potential for each spin component $\sigma=\uparrow,\downarrow$,
\begin{equation}
\bar{V}^{(H)}_{\sigma} = \frac{1}{N}\sum_i V^{(H)}_{i\sigma},
\label{HPaverage}
\end{equation}
where $N$ is the total number of lattice sites, the diagonal terms of the Hamiltonian in Eq.~(\ref{hamiltonian}) can be written as
\begin{equation}
\sum_{i\sigma} (\epsilon_i -\tilde{\mu} + \tilde{h}_\sigma + V_{i \bar{\sigma}}^{(H)}-\bar{V}_{\bar{\sigma}}^{(H)}) c^\dag_{i \sigma} c_{i \sigma}
\end{equation}
with $\tilde{h}_\sigma=-\tilde{h}$ ($+\tilde{h}$) for $\sigma=\uparrow$ ($\sigma=\downarrow$).
We have defined
\begin{eqnarray}
\tilde{\mu} &=& \mu-\frac{\bar{V}^{(H)}_{\up}+\bar{V}^{(H)}_{\down}}{2}\,,\label{shiftedmu}\\
\tilde{h} &=& h+\frac{\bar{V}^{(H)}_{\up}-\bar{V}^{(H)}_{\down}}{2}\,.\label{shiftedh}
\end{eqnarray}
For $h>0$, typically there are more spin-up electrons than spin-down electrons and the Hartree potential effectively reduces the Zeeman field. Intuitively, the average energy gain by the pairing interaction with spin-up electrons makes the electron to have its spin down less costly in terms of the Zeeman energy. When the system has translational symmetry, $V_{i \sigma}^{(H)}=\bar{V}^{(H)}_{\sigma};\forall i$, and the Hamiltonian (\ref{hamiltonian}) can be expressed in momentum space as
\begin{equation}
\mathcal{H} = \frac{1}{2} \sum_{\bm{k}}\Psi^\dagger_{\bm{k}} \mathcal{H}(\bm{k}) \Psi_{\bm{k}}\,,
\label{hkspace}
\end{equation}
where $\Psi_{\bm{k}}=(c_{\bm{k}\uparrow}\;c_{\bm{k}\downarrow}\;c^\dagger_{-\bm{k}\uparrow}\;c^\dagger_{-\bm{k}\downarrow})^T$ and
\begin{equation}
\mathcal{H}(\bm{k}) = 
\begin{pmatrix}
	\epsilon(\bm{k}) - \tilde{h}\sigma_z + \alpha \mathcal{L}(\bm{k})\cdot \bm{\sigma}  & i \Delta(\bm{k}) \sigma_y \\
	-i \Delta(\bm{k})^* \sigma_y & -\epsilon(\bm{k}) + \tilde{h}\sigma_z + \alpha  \mathcal{L}(\bm{k})\cdot \bm{\sigma}^* 
\end{pmatrix},
\label{hkmatrix}
\end{equation}
with $\epsilon(\bm{k}) = -2t({\rm cos}\,k_x + {\rm cos}\,k_y) -\tilde{\mu}$ and $\mathcal{L}(\bm{k})\equiv (\mathcal{L}_x, \mathcal{L}_y) = ({\rm sin}\,k_y, -{\rm sin}\,k_x)$. 
$c^\dagger_{\bm{k}\sigma}$ and $c_{\bm{k}\sigma}$ are the creation and annihilation operators of the electron with momentum $\bm{k}=(k_x,k_y)$ and spin $\sigma$, and $\bm{\sigma}\equiv (\sigma_x,\sigma_y)$ and $\sigma_z$ are the Pauli matrices. $\mathcal{H}(\bm{k})$ above reduces to the momentum-space Hamiltonian given in Ref.~\onlinecite{Sato2010} when the Hartree potential is neglected, with $\tilde{\mu}\equiv \mu$ and $\tilde{h}\equiv h$. Thus, various topological phases as classified in Ref.~\onlinecite{Sato2010} according to the first Chern number or the Thouless-Kohmoto-Nightingale-Nijs (TKNN) number\cite{TKNN1982} can be achieved by replacing the chemical potential and Zeeman field by $\tilde{\mu}$ and $\tilde{h}$, respectively, when the Hartree potential is taken into account.
Topological phase transitions between topologically distinct phases occur when the energy gap of the bulk quasiparticle spectrum closes.\cite{Sato2010} Assuming an isotropic $s$-wave order parameter, $\Delta(\bm{k})\equiv \Delta_0$, diagonalization of $\mathcal{H}(\bm{k})$ in Eq.~(\ref{hkmatrix}) yields
\begin{equation}
E_\pm(\bm{k}) = \sqrt{\epsilon(\bm{k})^2 + \alpha^2 |\mathcal{L}(\bm{k})|^2 + \tilde{h}^2 + |\Delta_0|^2 \pm 2\xi(\bm{k})}\,,
\label{bulkspectrum}
\end{equation}
where $\xi(\bm{k}) = \sqrt{\epsilon(\bm{k})^2 \alpha^2 |\mathcal{L}(\bm{k})|^2 + (\epsilon(\bm{k})^2 + |\Delta_0|^2)\tilde{h}^2}$, and the minimum value of $E_\pm(\bm{k})$ is the bulk spectral gap $E_0$. For example, as $\tilde{h}$ is varied for a given set of $\alpha$, $\tilde{\mu}$, and $\Delta_0$, the system transitions from one topological phase (trivial, Abelian, or non-Abelian) to another every time $E_0$ vanishes. The topological invariant that classifies each phase is the TKNN number\cite{TKNN1982} and can be calculated by\cite{Ishikawa1987,Volovik2009,Gurarie2011}
\begin{eqnarray}
\nu = &&\frac{1}{8\pi^2}\int d\bm{k} d\omega\, [\,{\rm Tr}(G\partial_{k_x}G^{-1}G\partial_{k_y}G^{-1}G\partial_{\omega}G^{-1})\nonumber\\
&&-{\rm Tr}(G\partial_{k_y}G^{-1}G\partial_{k_x}G^{-1}G\partial_{\omega}G^{-1})\,]\,,
\end{eqnarray}
where $G=(i\omega-\mathcal{H}(\bm{k}))^{-1}$ and $\nu\in\mathbb{Z}$.\cite{Schnyder2008}
The system is in trivial phase when $\nu=0$, and in Abelian and non-Abelian phase, respectively, when $\nu$ is even and odd ($-2$ and $\pm 1$ in this model). 

In this 2D $s$-wave TSC model, there are two underlying chiralities, $\eta_{\pm}\equiv -({\cal L}_x\pm i{\cal L}_y)/\sqrt{{\cal L}_x^2+{\cal L}_y^2}=\pm i({\rm sin}\,k_x \pm i\,{\rm sin}\,k_y)/\sqrt{{\rm sin}^2k_x+{\rm sin}^2k_y}$,\cite{Fujimoto2008,Sato2009p-wave} as can be seen by expressing $\mathcal{H}(\bm{k})$ in the ``chirality basis'' that diagonalizes the normal-state Hamiltonian $\epsilon(\bm{k}) - \tilde{h}\sigma_z + \alpha \mathcal{L}(\bm{k})\cdot \bm{\sigma}$:\cite{Sato2010,Smith2016} 
\begin{equation}
\tilde{\mathcal{H}}(\bm{k}) = 
\begin{pmatrix}
	\epsilon(\bm{k}) + \Delta\epsilon(\bm{k})\sigma_z & \hat{\Delta} \\
	\hat{\Delta}^\dagger & -\epsilon(\bm{k}) - \Delta\epsilon(\bm{k})\sigma_z
\end{pmatrix},
\label{hktilde}
\end{equation}
where $\Delta\epsilon(\bm{k}) = {\rm sgn}(\tilde{h}) \sqrt{ \alpha^2 |\mathcal{L}(\bm{k})|^2 + \tilde{h}^2}$ and
\begin{equation}
\hat{\Delta} = \frac{1}{\Delta\epsilon(\bm{k})}
\begin{pmatrix}
	-\alpha|\mathcal{L}(\bm{k})|\eta_+\Delta(\bm{k}) & \tilde{h}\Delta(\bm{k}) \\
	-\tilde{h}\Delta(\bm{k}) & -\alpha|\mathcal{L}(\bm{k})|\eta_-\Delta(\bm{k}) 
\end{pmatrix}.
\label{deltahat}
\end{equation}
The \emph{intraband} pairing in the band $E_+$ ($E_-$) has the chirality of $\eta_+$ ($\eta_-$), while the interband pairing is purely $s$-wave. This is analogous to the two chiralities $p_x\pm ip_y$ present in the nontrivial (non-Abelian) phase of the continuum model.\cite{Zhang2008,Alicea2010,Shitade2015} The two chiralities $\eta_{\pm}$ associated with the two Fermi surfaces are always mixed in Abelian phase. In contrast, the chirality associated with the single Fermi surface can be dominant over the other in non-Abelian phase for relatively weak SO coupling, and which chirality is more manifest is determined by the sign of $h$ as well as the sign of $\mu$.\cite{Sato2010,Wu2012,Smith2016} When SO coupling is strong, the two chiralities can be mixed strongly also in non-Abelian phase.\cite{Masaki2015,Smith2016}

To obtain the mean fields, we perform self-consistent iterations up to the $l$-th iteration step, where, e.g., the order parameter as a complex vector $\vec{\Delta}$ of length $N_xN_y$ satisfies
\begin{equation}
\frac{\|\vec{\Delta}^{(l)}-\vec{\Delta}^{(l-1)}\|}{\|\vec{\Delta}^{(l-1)}\|} < 10^{-6}
\end{equation}
and similarly for each spin component of the Hartree potential that can be regarded as a real vector of length $N_xN_y$. All the calculation presented below has been performed for zero temperature.

\section{\label{sec:results}Results}

\subsection{\label{sec:self}Self-consistency}

\begin{figure}
  \includegraphics[width=0.85\columnwidth]{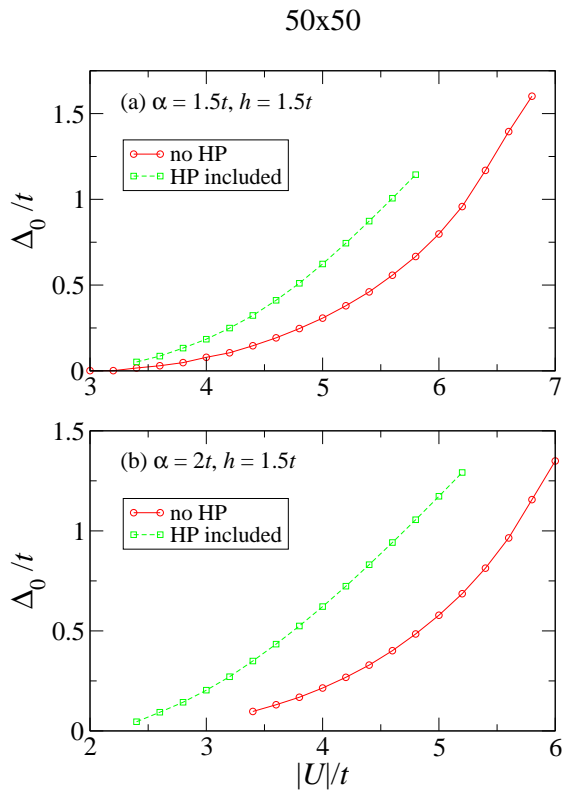}
  \vspace{.1cm}
  \caption{\label{fig:Delta_vs_U} (Colour online) (a) Order parameter $\Delta_0/t$ as a function of the pairing interaction strength $|U|/t$ in a uniform 50$\times$50 lattice with PBC for (a) $\alpha=1.5t$, $h=1.5t$ and (b) $\alpha=2t$, $h=1.5t$. The chemical potential without (with) the Hartree potential included is $\mu=-3t$ ($\tilde{\mu}=-3t$) and $\mu=t$ ($\tilde{\mu}=t$) for (a) and (b), respectively.
  }
\end{figure}

We first illustrate how the order parameter in a uniform system $\Delta_0\equiv \Delta_i;\forall i$ varies with the strength of the pairing interaction $|U|$. In our self-consistent calculation, the order parameter is defined in terms of Eq.~(\ref{order}) such that $\Delta_0$ is positive. Figure~\ref{fig:Delta_vs_U} presents the order parameter $\Delta_0$ as a function of $|U|$ both in units of $t$ in a uniform 50$\times$50-site lattice with the periodic boundary condition (PBC), with (green squares) and without (red circles) solving for the Hartree potential. Without (with) the Hartree potential, $\mu=-3t$ ($\tilde{\mu}=-3t$), $\alpha=1.5t$, $h=1.5t$ and  $\mu=t$ ($\tilde{\mu}=t$), $\alpha=2t$, $h=1.5t$ for the systems shown in Fig.~\ref{fig:Delta_vs_U}(a) and (b), respectively. When the Hartree potential is neglected, for $\mu=-3t$ ($\mu=t$) the non-Abelian (Abelian) TSC state with the TKNN number $\nu=1$ ($\nu=-2$) is realised according to the conditions,\cite{Sato2010}
\begin{eqnarray}
&&(4t-|\mu|)^2+\Delta_0^2<h^2<\mu^2+\Delta_0^2\,;\quad \nu = 1\,,\label{nu1}\\
&&\mu^2+\Delta_0^2<h^2<(4t-|\mu|)^2+\Delta_0^2\,;\quad \nu = -2\,,\label{nu-2}
\end{eqnarray}
and the system is in the trivial phase with $\nu=0$ when $(4t-|\mu|)^2+\Delta_0^2>h^2$ ($\mu^2+\Delta_0^2>h^2$). In both systems presented in Fig.~\ref{fig:Delta_vs_U} the transition to the trivial phase will occur for $\Delta_0 \gtrsim 1.12t$. This limits the range of the pairing interaction to $|U|<6.4t$ and $|U|<5.8t$, respectively, for the systems shown in Fig.~\ref{fig:Delta_vs_U}(a) and (b) to stay in the nontrivial phase.
As can be seen above, the phase boundaries do not depend on the signs of $\mu$ and $h$. The self-consistently obtained $\Delta_0$ is not affected by the sign of $\mu$ or $h$ either.

With the inclusion of the Hartree potential, $\mu$ and $h$ in Eqs.~(\ref{nu1}) and (\ref{nu-2}) are replaced by $\tilde{\mu}$ and $\tilde{h}$. For $\tilde{\mu}=-3t$, $\alpha=1.5t$, and $h=1.5t$, the system now transitions from the non-Abelian to trivial phase at around $|U|=4.8t$, where we have found $\tilde{h}\simeq 1.1t$ and $\Delta_0\simeq 0.5t$. The system with $\tilde{\mu}=t$, $h=1.5t$ and $\alpha=2t$ has a very narrow region of Abelian TSC for $|U|<3.2t$, where at $|U|=3.2t$ we have obtained $\tilde{h}=1.03t$ and $\Delta_0\simeq 0.27t$. Thus, in general, due to the effective reduction of the Zeeman field by the Hartree potential, one needs not assume too strong a coupling constant for the pairing and/or Rashba SO interaction when solving for the Hartree potential self-consistently. What this implies in terms of real materials is that the Pauli depairing effect of the Zeeman field is overcome partially by the Hartree potential.

\subsection{\label{sec:impurity}Effects of a single nonmagnetic impurity}

In a conventional $s$-wave superconductor, a nonmagnetic impurity can locally suppress the order parameter and cause Friedel-like oscillations around it;\cite{Tanaka2000,Balatsky2006} however, a single nonmagnetic impurity does not bind a quasiparticle. On the other hand, in a spin-triplet chiral $p$-wave superconductor, a nonmagnetic impurity can create midgap quasiparticle excitation and spontaneous supercurrent around it.\cite{Okuno2000} Furthermore, in the $p_x+ip_y$ domain where the $p_x+ip_y$ order is suppressed around a nonmagnetic impurity, the $p_x-ip_y$ order is induced, and vice versa.\cite{Okuno2000,Takigawa2005} It is an intriguing question as to how nonmagnetic impurities affect 2D $s$-wave TSC due to the presence of the two underlying chiralities $\eta_\pm$ ($p_x\pm ip_y$ in the continuum model).\cite{Hu2013,Nagai2014oval,Nagai2015,Shitade2015}

Nagai, Ota, and Machida have examined the average effects of nonmagnetic impurities in the 2D $s$-wave TSC model\cite{Sato2009,Sato2010} by solving for the impurity self-energy and the uniform order parameter self-consistently.\cite{Nagai2014oval} They have shown that while the system in trivial phase ($\nu=0$) is robust against nonmagnetic impurities, the Anderson theorem\cite{Anderson1959} can break down in a nontrivial phase ($\nu\ne 0$) and that the superconducting transition temperature $T_c$ becoming sensitive to impurity concentration is accompanied by the appearance of midgap states. They have further studied quasiparticle bound states around a single nonmagnetic impurity in non-Abelian phase in terms of the BdG equations (without self-consistency)\cite{Nagai2015} and have found that the bound-state energy decreases with respect to the bulk spectral gap as the Zeeman field $h$ is increased. This is consistent with their finding of $T_c$ being more sensitive to nonmagnetic impurities for stronger Zeeman field in Ref.~\onlinecite{Nagai2014oval} and with the picture that the intraband chiral $p$-wave pairing (with chirality $\eta_+$ or $\eta_-$ in non-Abelian phase) dominates over the interband $s$-wave pairing for large $|h|$.\cite{Alicea2010,Nagai2015}

Hu {\it et al}.\cite{Hu2013} have obtained ``universal'' midgap bound states around a nonmagnetic point (delta-function) impurity in a 2D $s$-wave superfluid with Rashba SO coupling and Zeeman field in a harmonic trap by solving the radial BdG equations self-consistently. They have found for one set of parameters in non-Abelian phase that a strong point impurity (nonmagnetic or magnetic) results in a midgap state with the ``universal'' bound-state energy $\sim \Delta_0^2/E_F$, where $\Delta_0$ and $E_F$ are the bulk order parameter and the Fermi energy, respectively. As pointed out by Shitade and Nagai,\cite{Shitade2015} this bound-state energy is not universal as it was derived by substituting the $s$-wave order parameter $\Delta_0$ in the formula for Majorana edge modes in a chiral $p$-wave superfluid in confinement.\cite{Stone2004} The argument made by the authors of Ref.~\onlinecite{Hu2013} is that the impurity site where the order parameter is suppressed to zero acts as a singular point of vacuum and its ``interface'' with the nontrivial region hosts a Majorana edge state. We note, however, that the self-consistently solved order parameter increases continuously from zero away from the impurity site and there is no clear boundary with which trivial and nontrivial regions can be defined locally.

\begin{figure}
  \includegraphics[width=0.85\columnwidth]{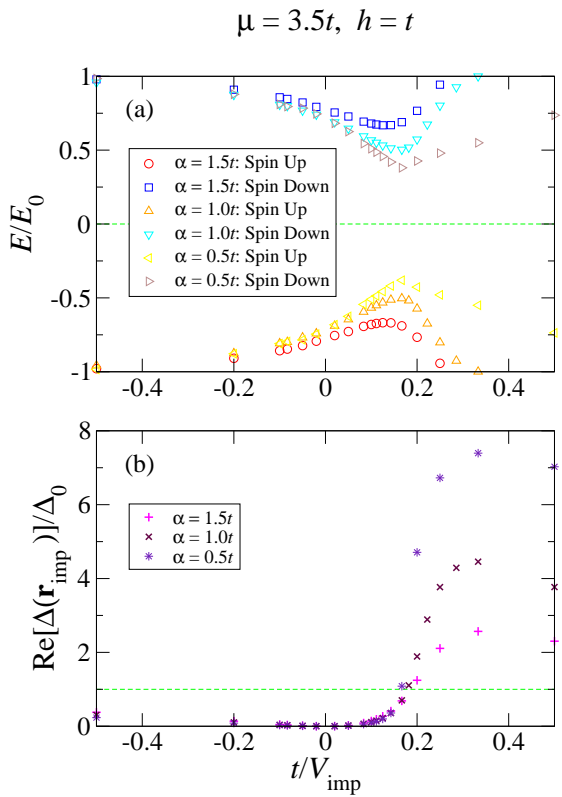}
  \caption{\label{fig:Vimp_mu3.5_h1} (Colour online) (a) Energy of the quasiparticle excitation bound to the nonmagnetic impurity scaled to the bulk spectral gap $E_0$ and (b) the real part of the order parameter at the impurity site in units of the bulk order parameter $\Delta_0$ as a function of the inverse of the impurity potential, $t/V_{\rm imp}$, for $\mu=3.5t$ and $h=t$; for $\alpha=1.5t$, $t$ and $0.5t$. $\Delta_0\simeq 0.34t$ and $\nu=1$.
  }
\end{figure}

\begin{figure}
  \includegraphics[width=0.85\columnwidth]{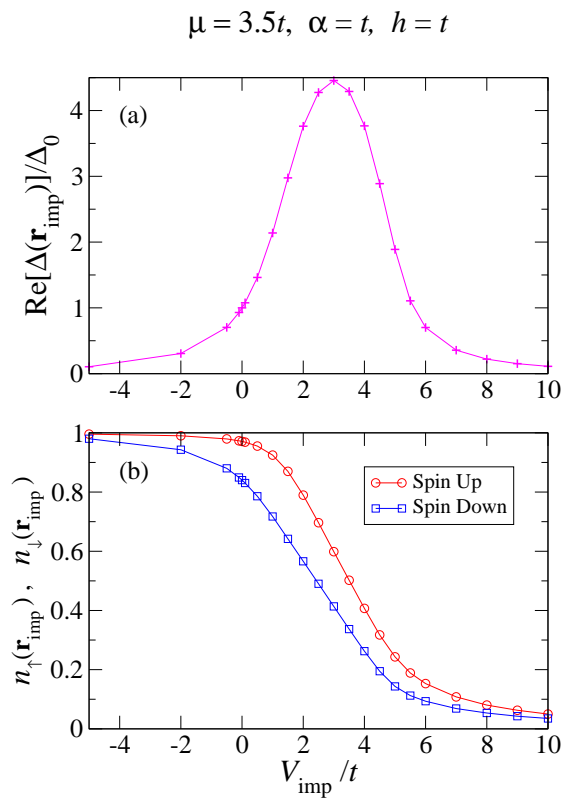}
  \caption{\label{fig:Vimp_mu3.5_h1_alpha1} (Colour online) (a) Real part of the order parameter in units of the bulk order parameter $\Delta_0$ and (b) the average number of spin-up and spin-down electrons at the impurity site as a function of the impurity potential, $V_{\rm imp}/t$, for $\mu=3.5t$, $h=t$, and $\alpha=t$ in a 64$\times$64 lattice.
  }
\end{figure}

In this subsection, we demonstrate that the effects of a nonmagnetic impurity on 2D $s$-wave TSC are far from being ``universal'' and depend significantly on the material parameters ($\mu$, $\alpha$, and $h$). For relatively weak SO coupling in non-Abelian phase, where one of the two chiralities $\eta_\pm$ can be dominant over the other, we find in general that the smaller the $\alpha$, or the larger the $|h|$, the more $p$-wave-like the system reacts to a nonmagnetic impurity. Reversing the sign of $h$ while keeping all the other parameters the same does not change the order parameter nor the excitation spectrum: only the spin components of quasiparticle excitation are interchanged. In case where one chirality is dominant over the other in non-Abelian phase, $h\rightarrow -h$ or $\mu\rightarrow -\mu$ switches the major chirality.\cite{Wu2012,Sato2010,Smith2016} Under $\mu\rightarrow -\mu$, changing the sign of the impurity potential $V_{\rm imp}$ at the same time leaves the order parameter and excitation spectrum virtually unchanged. In the results presented below, $h>0$ and the Hartree potential was neglected.

We first illustrate the effects of the SO coupling strength $\alpha$. We place a single nonmagnetic impurity with potential $V_{\rm imp}$ at the centre of the lattice with PBC. The order parameter acquires a small imaginary part in the vicinity of a nonmagnetic impurity, while it remains real at the impurity site. In Fig.~\ref{fig:Vimp_mu3.5_h1} we show (a) the quasiparticle bound-state energy in units of the bulk spectral gap $E_0$ and (b) the ratio of the real part of the order parameter at the impurity site $\bm{r}_{\rm imp}$ to the bulk order parameter $\Delta_0$ as a function of $t/V_{\rm imp}$ for $\mu=3.5t$ and $h=t$, for $\alpha=1.5t$, $t$, and $0.5t$. The system size is 51$\times$51 lattice sites for $\alpha=0.5t$ and $1.5t$, and 64$\times$64 for $\alpha=t$. We find that typically the disturbance of the order parameter by the impurity is contained well within 51$\times$51 lattice sites. The coupling constant for the pairing interaction has been chosen to be $U=-4.61t$, $-5.2t$ and $-6.24t$ for $\alpha=1.5t$, $t$ and $0.5t$, respectively, such that $\Delta_0\simeq 0.34t$ ($\nu=1$). The spin of the quasiparticle excitation has been determined by identifying the peak at the eigenvalue in the spin-resolved LDOS at the impurity site. Consistently with the non-self-consistent results of Ref.~\onlinecite{Nagai2015}, we find that the midgap excitation is always spin-polarised at the impurity site. 

\begin{figure}
  \includegraphics[width=0.95\columnwidth]{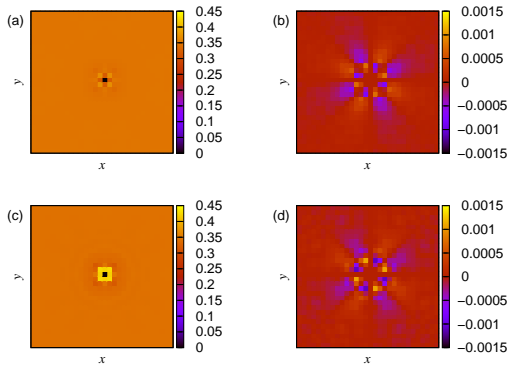}
  \caption{\label{fig:order_mu3.5_h1_Vimp-50} (Colour online) Real (a),(c) and imaginary (b),(d) parts of the order parameter $\Delta(x,y)/t$ as a function of $x$ and $y$ for $11\le x,y \le 41$ with $V_{\rm imp}=-50t$ at the centre of a 51$\times$51 lattice for $\mu=3.5t$ and $h=t$; for $\alpha=t$ (a),(b) and $\alpha=0.5t$ (c),(d).
  }
\end{figure}

\begin{figure}
  \includegraphics[width=0.85\columnwidth]{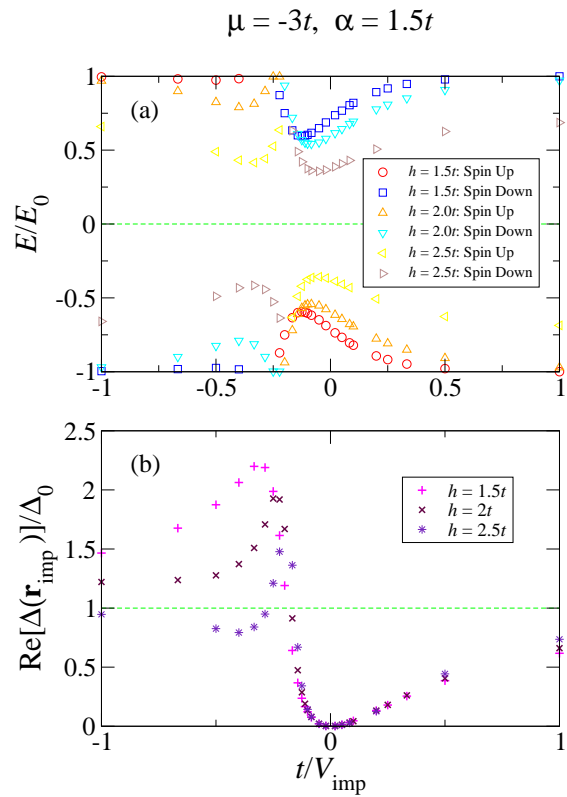}
  \caption{\label{fig:Vimp_mu-3_alpha1.5} (Colour online) (a) Quasiparticle bound-state energy $E/E_0$ and (b) ${\rm Re}[\Delta(\bm{r}_{\rm imp})]/\Delta_0$ as a function of $t/V_{\rm imp}$ for $\mu=-3t$ and $\alpha=1.5t$, for $h=1.5t$, $2t$ and $2.5t$ in a 51$\times$51 lattice. $\Delta_0\simeq 0.34t$ and $\nu=1$.
  }
\end{figure}

In these non-Abelian systems the chirality $\eta_+$ is dominant over $\eta_-$, as can be apparent in the core of a vortex pinned by a nonmagnetic impurity.\cite{Smith2016} One can see in Fig.~\ref{fig:Vimp_mu3.5_h1}(a) that the nonmagnetic impurity binds spin-down quasiparticle excitation (i.e., composed of spin-down particle and spin-up hole), which becomes more strongly bound to the impurity as $\alpha$ decreases when $V_{\rm imp}>0$. This can be understood by the fact that for a given $h$, the smaller the $\alpha$, the more manifest the major chirality becomes and the more chiral $p$-wave-like the system behaves.\cite{Shitade2015,Masaki2015,Smith2016} For $V_{\rm imp}<0$ the bound-state energy is similar and ${\rm Re}\,\Delta(x,y)$ is suppressed at the impurity site to a similar value for the three values of $\alpha$. In the range $V_{\rm imp}>0$ where the bound-state energy is substantially different for $\alpha=0.5t$, $t$, and $1.5t$, ${\rm Re}\,\Delta(x,y)$ is peaked at the impurity site, more sharply for smaller $\alpha$, as can be seen in Fig.~\ref{fig:Vimp_mu3.5_h1}(b). This is illustrated further in Fig.~\ref{fig:Vimp_mu3.5_h1_alpha1} for $\alpha=t$, where (a) ${\rm Re}\,\Delta(\bm{r}_{\rm imp})$ and (b) the average number of spin-up and spin-down electrons at the impurity site are plotted as a function of $V_{\rm imp}$. The average number of electrons decreases monotonically as $V_{\rm imp}$ increases, while ${\rm Re}\,\Delta(\bm{r}_{\rm imp})$ reaches its maximum value for  $V_{\rm imp}\simeq 3t$ in this system. 

The enhancement of the order parameter at the impurity site can be understood roughly by the fact that the chemical potential is shifted to $\mu-V_{\rm imp}$ locally and this effectively makes the site transition into the trivial phase. Although the phase boundaries that separate trivial and nontrivial phases are defined in terms of bulk $\mu$, $h$, and $\Delta_0$, and do not directly apply to a single lattice site, we have checked that with no impurity and the other parameters fixed, e.g., for $\mu=0.5t$, the BdG equations converge to a trivial state with uniform order parameter similar in value to ${\rm Re}\,\Delta(\bm{r}_{\rm imp})$ for $V_{\rm imp}=3t$ shown in Fig.~\ref{fig:Vimp_mu3.5_h1}(b), for $\alpha=t$ and $0.5t$.

In Fig.~\ref{fig:order_mu3.5_h1_Vimp-50} the order parameter $\Delta(x,y)/t$ is plotted as a function of spatial coordinates $x$ and $y$ for $11\le x,y \le 41$ with $V_{\rm imp}=-50t$ at the centre of a 51$\times$51-site system for $\mu=3.5t$ and $h=t$, for $\alpha=t$ (upper graphs) and $\alpha=0.5t$ (lower graphs). The left (right) panel shows the real (imaginary) part of the order parameter. This figure illustrates the chiral $p$-wave nature of the system being enhanced for smaller $\alpha$. As can be seen in Fig.~\ref{fig:order_mu3.5_h1_Vimp-50}, the order parameter is suppressed to nearly zero at the impurity site for both $\alpha=t$ and $\alpha=0.5t$. For $\alpha=0.5t$, however, the real part of the order parameter is enhanced at the nearest- and next-nearest-neighbour sites to the impurity (${\rm Re}\,\Delta(x,y)\approx 0.4t$). This is reminiscent of the order parameter of the opposite chirality induced around a nonmagnetic impurity in a chiral $p$-wave superconductor.\cite{Okuno2000,Takigawa2005} Moreover, the small, oscillatory imaginary part in the order parameter indicates the presence of induced supercurrent around the impurity, and the oscillation amplitudes are increased slightly for $\alpha=0.5t$ compared to those for $\alpha=t$. 

\begin{figure}
  \includegraphics[width=0.85\columnwidth]{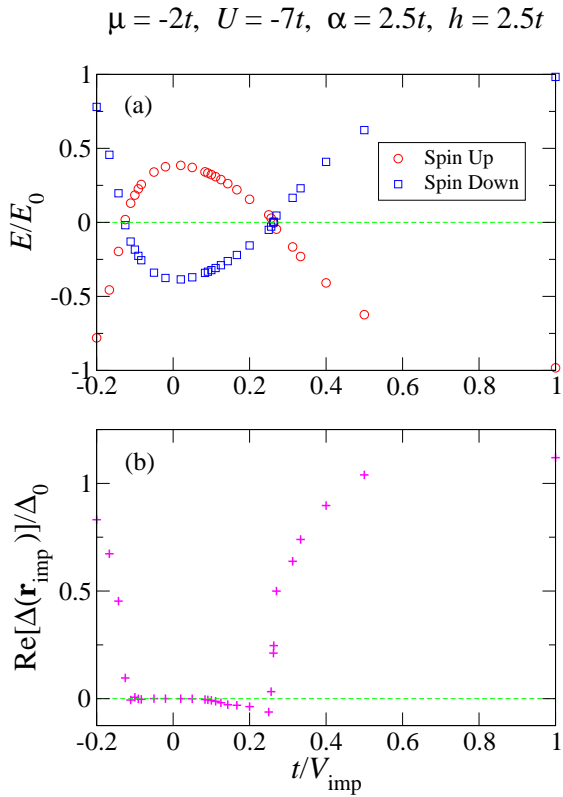}
  \caption{\label{fig:Vimp_mu-2_U-7} (Colour online) (a) Quasiparticle bound-state energy $E/E_0$ and (b) ${\rm Re}[\Delta(\bm{r}_{\rm imp})]/\Delta_0$ as a function of $t/V_{\rm imp}$ for $\mu=-2t$, $h=2.5t$, and $\alpha=2.5t$ in a 51$\times$51 lattice. $\Delta_0\simeq 0.34t$ and $\nu=-1$. 
  }
\end{figure}

\begin{figure}
  \includegraphics[width=0.95\columnwidth]{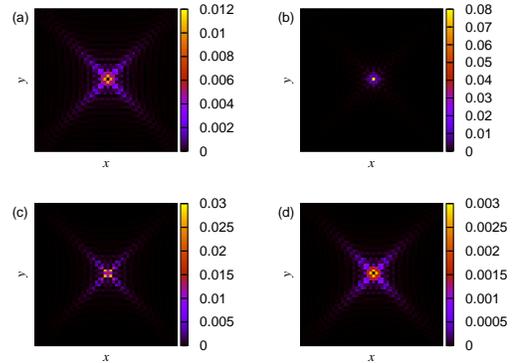}
  \caption{\label{fig:ldos_mu-2} (Colour online) Spin-up (a),(c) and spin-down (b),(d) components of the LDOS at the bound-state energy as a function of $x$ and $y$ for the entire 51$\times$51 lattice for $\mu=-2t$, $h=2.5t$ and $\alpha=2.5t$; with $V_{\rm imp}=-5t$ (a),(b) and $V_{\rm imp}=6t$ (c),(d) at the centre of the lattice.
  }
\end{figure}

We have found curious effects of a nonmagnetic impurity by increasing the Zeeman field $h$: the spin of the impurity bound state flips as the impurity potential is varied. Two different examples are presented below. Shown in Fig.~\ref{fig:Vimp_mu-3_alpha1.5} are (a) the quasiparticle bound-state energy $E/E_0$ and (b) the order parameter at the impurity site ${\rm Re}\,\Delta(\bm{r}_{\rm imp})/\Delta_0$ as a function of $t/V_{\rm imp}$ for $\mu=-3t$ and $\alpha=1.5t$ on a 51$\times$51 lattice, for $h=1.5t$, $2t$, and $2.5t$. $U=-5.1t$, $-6.02t$ and $-7.19t$ for $h=1.5t$, $2t$ and $2.5t$, respectively, that yield $\Delta_0\simeq 0.34t$ ($\nu=1$). The energy of the spin-down bound state varies in a way similar to the results shown in Fig.~\ref{fig:Vimp_mu3.5_h1}(a) as $t/V_{\rm imp}$ is decreased from the positive side, reaching a maximum at some negative value of $V_{\rm imp}$. The bound state, however, disappears into the gap edge rather abruptly as $t/V_{\rm imp}$ is decreased further and then reappears with spin up; although for $h=1.5t$ it remains very close to the gap edge. In this region the order parameter is peaked at the impurity site, once again due to the local shift of the chemical potential to $\mu-V_{\rm imp}$ that effectively transforms the site into another phase -- either trivial or Abelian depending on the value of $V_{\rm imp}$.

Figure~\ref{fig:Vimp_mu-2_U-7} demonstrates a striking example where a nonmagnetic impurity in a 2D $s$-wave topological superconductor acts like a magnetic impurity (classical spin) in a conventional $s$-wave superconductor. The system presented in this figure is a 51$\times$51 lattice with $\mu=-2t$, $h=2.5t$, $\alpha=2.5t$, $U=-7t$ and $\Delta_0\simeq 0.34t$ ($\nu=-1$). A magnetic impurity in a conventional $s$-wave superconductor locally breaks time-reversal symmetry and creates spin-polarised quasiparticle excitation (the spin direction can be defined with respect to that of the impurity) in the energy gap\cite{Yu1965,Shiba1968,Rusinov1969} and as the potential strength of the impurity is increased, the excitation energy crosses zero, after which the ground state contains a spin-polarised quasiparticle (spin-$1/2$ up or down, depending on the definition) bound to the impurity and the midgap excitation has the opposite spin as it would now remove this extra spin from the ground state.\cite{Sakurai1970} When this phase transition of the ground state happens, the order parameter becomes negative at the impurity site\cite{Salkola1997,Flatte1997l} -- caused by the sign change in the contribution from the impurity bound state\cite{Flatte1997l,Flatte1997b} -- and vanishes in the limit of infinite potential strength.\cite{Salkola1997}

It can be seen in Fig.~\ref{fig:Vimp_mu-2_U-7}(a) that as $V_{\rm imp}$ is increased in magnitude (either from the positive or negative side), the energy of the spin-down quasiparticle excitation decreases and crosses zero, at which point the quasiparticle excitation becomes spin up and ${\rm Re}\,\Delta(\bm{r}_{\rm imp})$ changes sign from positive to negative, as can be seen in Fig.~\ref{fig:Vimp_mu-2_U-7}(b). As $|V_{\rm imp}|$ is increased further, ${\rm Re}\,\Delta(\bm{r}_{\rm imp})$ approaches zero. Since the two limits $t/V_{\rm imp}\rightarrow \pm\infty$ ($V_{\rm imp}\rightarrow 0\pm$) should result in the same uniform state, the impurity bound-state energy must cross zero an even number of times:\cite{Sau2013} this is indeed the case for the quasiparticle excitation shown in Fig.~\ref{fig:Vimp_mu-2_U-7}(a). The zero-energy crossing occurs at $V_{\rm imp}\approx -8t$ and $V_{\rm imp}\approx 3.8t$ in this system. We have found the excitation (spin up at the impurity site) with energy $E\simeq 2.6\times 10^{-5}t$ for $V_{\rm imp}=3.8125t$, which is \emph{not} a Majorana fermion.
The spin-resolved LDOS at the bound-state energy is plotted as a function of $x$ and $y$ coordinates for the entire lattice in Fig.~\ref{fig:ldos_mu-2} for $V_{\rm imp}=-5t$ (upper panel) and $V_{\rm imp}=6t$ (lower panel); where the spin of the quasiparticle excitation is down and up, respectively, at the impurity site. As can be seen in Fig.~\ref{fig:ldos_mu-2}, the spin-up (spin-down) LDOS is zero at the impurity site for $V_{\rm imp}=-5t$ ($V_{\rm imp}=6t$). The diagonal extension of the LDOS reflects $k_y=\pm k_x$ in the Fermi wave vector on large portions of the squarish Fermi surface.\cite{Smith2016} The LDOS was calculated by using the eigenfunctions obtained by the SS method.\cite{Sakurai2003,Nagai2013}

\begin{figure}
  \includegraphics[width=0.9\columnwidth]{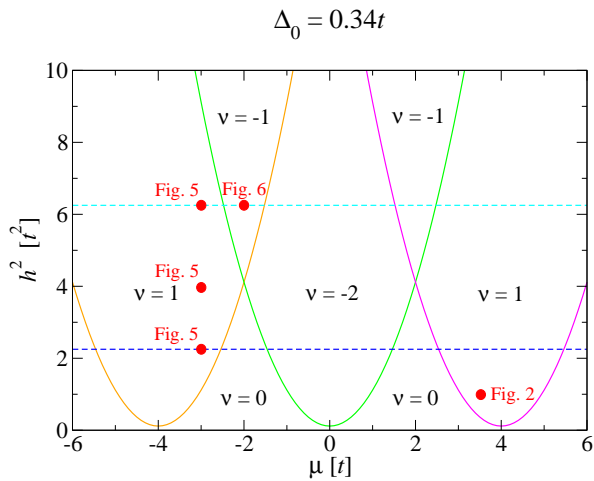}
  \caption{\label{fig:phase} (Colour online) Phase diagram of the topological phases with $\nu=0$ (trivial), $\nu=\pm 1$ (non-Abelian) and $\nu=-2$ (Abelian), separated by the phase boundaries, $h^2=(4t+\mu)^2+\Delta_0^2$, $h^2=\mu^2+\Delta_0^2$ and $h^2=(4t-\mu)^2+\Delta_0^2$ as a function of $h$ and $\mu$ with $\Delta_0=0.34t$.\cite{Sato2010} Two values of the Zeeman field, $h=1.5t$ and $h=2.5t$, are indicated by dashed lines. 
  }
\end{figure}

We have also found the same phase transition of the ground state in the presence of a nonmagnetic impurity in the system with $\mu=-2t$, $h=2.5t$, $\alpha=1.5t$, $U=-8.49t$ and $\Delta_0\simeq 0.34t$ ($\nu=-1$), where the variation of the bound-state energy as a function of $t/V_{\rm imp}$ is very similar to that shown in Fig.~\ref{fig:Vimp_mu-2_U-7}(a) for $V_{\rm imp}>0$, but the zero crossing on the negative side occurs at $V_{\rm imp}\approx -4.7t$, in contrast to $V_{\rm imp}\approx -8t$ for the system shown in Fig.~\ref{fig:Vimp_mu-2_U-7} ($\alpha=2.5t$). It is interesting to note that for $h=2.5t$, $\alpha=1.5t$, and $\Delta_0\simeq 0.34t$, the spin of the impurity bound state flips as the impurity potential is varied with and without crossing zero energy, respectively, for $\mu=-2t$ ($\nu=-1$) and $\mu=-3t$ ($\nu=1$) [Fig.~\ref{fig:Vimp_mu-3_alpha1.5}(a)].

We summarise the parameter values used for Figs.~\ref{fig:Vimp_mu3.5_h1}, \ref{fig:Vimp_mu-3_alpha1.5} and \ref{fig:Vimp_mu-2_U-7} in the phase diagram as a function of $h$ and $\mu$ shown in Fig.~\ref{fig:phase}, where different topological phases with $\nu=0$, $\nu=\pm 1$, and $\nu=-2$ are separated by the phase boundaries, $h^2=(4t\pm\mu)^2+\Delta_0^2$ and $h^2=\mu^2+\Delta_0^2$, with $\Delta_0=0.34t$.\cite{Sato2010}
The system is in the trivial phase with $\nu=0$ when $h^2<(4t\pm\mu)^2+\Delta_0^2$ and $h^2<\mu^2+\Delta_0^2$. 
Although there is no direct correspondence of the phase diagram to the single impurity site, for the systems presented in Fig.~\ref{fig:Vimp_mu3.5_h1} and for $h=1.5t$ in Fig.~\ref{fig:Vimp_mu-3_alpha1.5}, the range of $V_{\rm imp}$ where the order parameter is peaked at the impurity site loosely corresponds to ranges of the local chemical potential $\mu-V_{\rm imp}$ where $\nu=0$ or $-2$. In contrast, for $h=2.5t$ in Fig.~\ref{fig:Vimp_mu-3_alpha1.5} ($\nu=1$) and for the system shown in Fig.~\ref{fig:Vimp_mu-2_U-7} ($\nu=-1$), the TKNN number associated with the local chemical potential varies among $\nu=1$, $-1$ and $-2$ and reaches $\nu=0$ only for $|V_{\rm imp}/t|\gg 1$. Interestingly, the spin reversal of the midgap excitation occurs for $\mu-V_{\rm imp}\sim 1.5t$ ($V_{\rm imp}\sim -4.5t$) for $h=2.5t$ in Fig.~\ref{fig:Vimp_mu-3_alpha1.5} and for $\mu-V_{\rm imp}\approx 6t$ and $-5.8t$ ($V_{\rm imp}\approx -8t$ and $V_{\rm imp}\approx 3.8t$) for the system shown in Fig.~\ref{fig:Vimp_mu-2_U-7}, which correspond to $\nu=-1$ and $\nu=1$, respectively. Namely, the ``local'' TKNN number at the impurity site has the opposite sign to that of the bulk TKNN number, which is analogous to locally changing the sign of $h$\cite{Smith2016} and thus flipping the spin of the quasiparticle excitation. This does not explain, however, the rather striking difference in the behaviour of the impurity bound state between Figs.~\ref{fig:Vimp_mu-3_alpha1.5} and \ref{fig:Vimp_mu-2_U-7}, which warrants further investigation.

Sau and Demler have found similar zero-energy crossing of midgap excitation bound to a nonmagnetic impurity in a 1D semiconductor with Rashba SO coupling and Zeeman field in proximity to an $s$-wave superconductor.\cite{Sau2013} In this system, the bound-state energy also crosses zero in the limit of infinitely strong impurity potential, as it corresponds to the system effectively cut in half by the impurity and the zero-energy states are the Majorana fermions bound at the two edges. Thus, in the case of a 1D topological superconductor, there is an odd number of zero-energy crossings for finite impurity potential. 

\subsection{\label{sec:cdw}Coexistence of TCDW and TSC}

\begin{figure}
  \includegraphics[width=0.85\columnwidth]{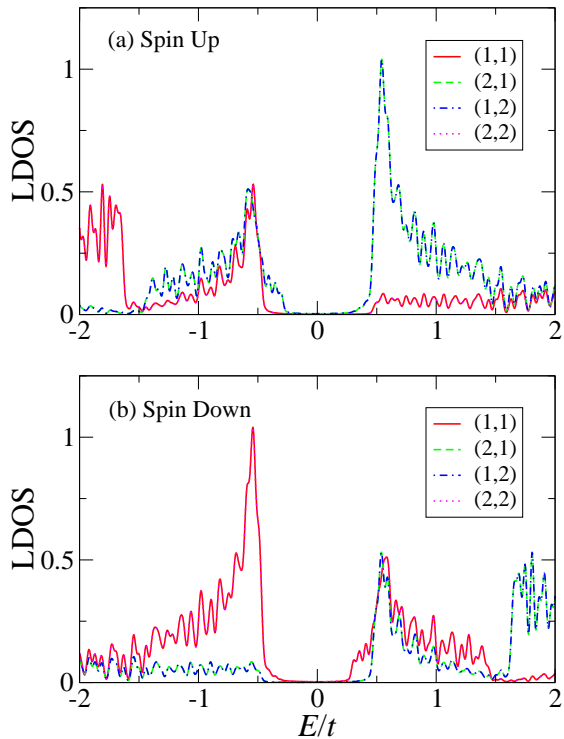}
  \vspace{.1cm}
  \caption{\label{fig:TCDW_LDOS} (Colour online) (a) Spin-up and (b) spin-down components of the LDOS as a function of excitation energy $E/t$ at lattice sites $(x,y)$ = (1,1), (2,1), (1,2), and (2,2) in the pure TCDW state for $\alpha=t$, $h=1.5t$, and $U=-4t$ in a 64$\times$64 lattice with PBC.
  }
\end{figure}

The 2D s-wave TSC model can also host topological charge density waves (TCDW).\cite{Sato2010} With the order parameter $\Delta_{\rm C}\equiv \langle c^\dagger_{\bm{k}\uparrow}c_{\bm{k}+\bm{Q}\uparrow} \rangle = \langle c^\dagger_{\bm{k}\downarrow}c_{\bm{k}+\bm{Q}\downarrow} \rangle$ for charge density waves (CDW) with wavevector $\bm{Q}$, the mean-field Hamiltonian can be written as\cite{Sato2010}
\begin{equation}
\mathcal{H}_{\rm CDW} = \frac{1}{2} \sum_{\bm{k}}\Psi^\dagger_{{\rm C}\bm{k}} \mathcal{H}_{\rm CDW}(\bm{k}) \Psi_{{\rm C}\bm{k}}\,,
\label{hcdw}
\end{equation}
where $\Psi_{{\rm C}\bm{k}}=(c_{\bm{k}\uparrow}\;c_{\bm{k}\downarrow}\;-c_{\bm{k}+\bm{Q}\downarrow}\;c_{\bm{k}+\bm{Q}\uparrow})^T$ and $\mathcal{H}_{\rm CDW}(\bm{k}) =$
\begin{equation}
\begin{pmatrix}
	\epsilon(\bm{k}) - \tilde{h}\sigma_z + \alpha \mathcal{L}(\bm{k})\cdot \bm{\sigma}  & i \Delta_{\rm C} \sigma_y \\
	-i \Delta_{\rm C} \sigma_y & \epsilon(\bm{k}+\bm{Q}) + \tilde{h}\sigma_z - \alpha  \mathcal{L}(\bm{k}+\bm{Q})\cdot \bm{\sigma}^* 
\end{pmatrix}.
\label{cdwmatrix}
\end{equation}
We note that the $\mathcal{L}(\bm{k}+\bm{Q})$ term in $\mathcal{H}_{\rm CDW}(\bm{k})$ given in Ref.~\onlinecite{Sato2010} ($\tilde{h}\equiv h$) has a wrong sign: it should be minus as above. If $\bm{Q}$ satisfies $\epsilon(\bm{k}+\bm{Q})=-\epsilon(\bm{k})$ and $\mathcal{L}(\bm{k}+\bm{Q})=-\mathcal{L}(\bm{k})$, $\mathcal{H}_{\rm CDW}(\bm{k})$ is equivalent to $\mathcal{H}(\bm{k})$ in Eq.~(\ref{hkmatrix}) for TSC. In our tight-binding model, these two conditions are satisfied for $\bm{Q}=(\pm\pi,\pm\pi)$ and $\tilde{\mu}=0$.
For $\tilde{\mu}=0$, the only nontrivial phase possible is Abelian with $\nu=-2$. For the same $\tilde{h}$, $\alpha$, and the uniform order parameter $\Delta_{\rm C}\equiv \Delta_0$, the TCDW and TSC states then have the same quasiparticle spectrum, with two zero-energy bound states per surface. In the TCDW state, however, the two zero-energy surface states are equivalent to each other due to folding of the Brillouin zone by the CDW order and hence there is only one zero mode per surface.\cite{Sato2010} Furthermore, while the zero modes are each a Majorana fermion in the TSC state, quasiparticle excitation in the TCDW state is not and possesses $U$(1) charge.\cite{Sato2010}

\begin{figure}
  \includegraphics[width=0.85\columnwidth]{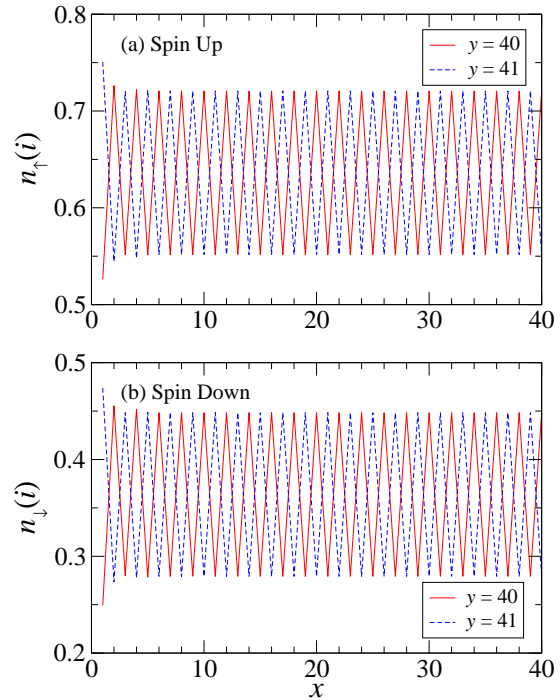}
  \vspace{.2cm}
  \caption{\label{fig:TCDW_xOBC} (Colour online) (a) Spin-up and (b) spin-down electron density as a function of $x$ at $y=40$ and $y=41$ in the TCDW state for $\alpha=2.5t$, $h=1.5t$, and $U=-4t$ on a 80$\times$80 lattice with open boundaries at $x=1$ and $x=80$.
  }
\end{figure}

We have solved the BdG equations with the Hamiltonian in Eq.~(\ref{hamiltonian}) self-consistently for each spin component of the Hartree potential with PBC, no impurity, and the superconducting (SC) order parameter set to zero, and have found uniform TCDW states with $\bm{Q}=(\pm\pi,\pm\pi)$ and $\tilde{\mu}=0$. For example, $\alpha=t$, $h=1.5t$, and $U=-4t$ on a 64$\times$64 lattice yield the TCDW state with $\Delta_{\rm C}=0.67504t$, the effective Zeeman field $\tilde{h}=0.96135t$, and $\nu=-2$. In real space, the CDW order parameter is the deviation of the Hartree potential from its average value, which alternates between $+\Delta_{\rm C}$ and $-\Delta_{\rm C}$ from one site to its nearest-neighbour site, in each spin component. The average electron density (number of electrons per site) is exactly 1.0. This confirms the fact that $\tilde{\mu}=0$ in the presence of Rashba SO coupling and Zeeman field corresponds to half filling, as for the tight-binding system for conventional $s$-wave superconductivity (with nearest-neighbour hopping only). 

We show in Fig.~\ref{fig:TCDW_LDOS} the (a) spin-up and (b) spin-down components of the LDOS as a function of excitation energy at lattice sites $(x,y)$ = (1,1), (2,1), (1,2), and (2,2) in the above TCDW state for $\alpha=t$, $h=1.5t$, and $U=-4t$ on a 64$\times$64 lattice. The LDOS was calculated using the Lorentz kernel\cite{Weisse2006,Covaci2010} with the corresponding Lorentzian smoothing width of $0.0005t$. The CDW order is such that the electron density $n_\uparrow(i)=0.803424$ (0.465904) and $n_\downarrow(i)=0.534096$ (0.196576) at sites (1,1) and (2,2) ((2,1) and (1,2)). These density modulations are reflected in the LDOS, showing more hole and electron excitations available for (1,1) and (2,1), respectively, for both spin components.

\begin{figure}
  \includegraphics[width=0.9\columnwidth]{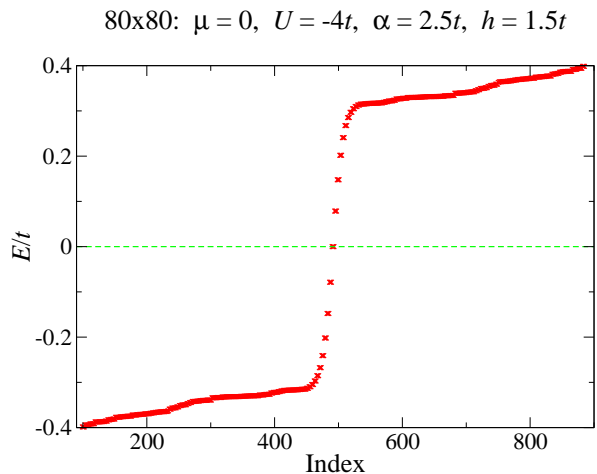}
  \caption{\label{fig:TCDW_eigs} (Colour online) Excitation spectrum in the TCDW state for $\alpha=2.5t$, $h=1.5t$, and $U=-4t$ on a 80$\times$80 lattice with open boundaries at $x=1$ and $x=80$. The index numbers the eigenvalues.
  }
\end{figure}

By solving for the SC order parameter self-consistently along with the Hartree potential for the same set of input parameters, we have also obtained the TSC state with the uniform electron density of 1.0 with $\Delta_0=0.67504t$ ($\Delta_{\rm C}=0$) and a mixed TSC+TCDW state with $\Delta_0=0.42169t$ and $\Delta_{\rm C}=0.52712t$, both with $\tilde{h}=0.96135t$. The three (pure TCDW, pure TSC, and mixed) states have exactly the same ground-state energy and excitation spectrum with the spectral gap $E_0\simeq 0.247t$. Thus, the TSC and TCDW states are degenerate ground states at half filling, as in the attractive Hubbard model for conventional $s$-wave superconductivity.\cite{Tanaka2000} Hence any linear superposition of the TSC and TCDW states is also a ground state with $\nu=-2$ and the two topological orders can coexist.

By introducing surface boundaries, we have also confirmed the existence of zero-energy bound states in our self-consistent solutions. We have obtained the pure TCDW state at $\tilde{\mu}=0$ for $\alpha=2.5t$, $h=1.5t$, and $U=-4t$ on a 80$\times$80 lattice with surface edges at $x=1$ and $x=80$ and PBC in the $y$ direction. A cross section of the (a) spin-up and (b) spin-down electron density is plotted as a function of $x\le 40$ at $y=40$ and $y=41$ in Fig.~\ref{fig:TCDW_xOBC}. The presence of an edge affects only few lattice sites close to the edge and we observe perfect $(\pm\pi,\pm\pi)$ CDW in the bulk of the system, where $\Delta_{\rm C}\simeq 0.338t$. The effective Zeeman field is $\tilde{h}=0.9552t$. The excitation spectrum is presented in Fig.~\ref{fig:TCDW_eigs}, where the abscissa is an index numbering the eigenvalues, clearly showing the existence of zero-energy states ($E\simeq 10^{-6}t$). We have also found the pure TSC state in which the SC order parameter is enhanced significantly at and very near $x=1$ and $x=80$, but with $\Delta_0\simeq 0.338t$ in the bulk and two Majorana bound states per surface with $E\simeq 10^{-6}t$. Moreover, a mixed state has been found where the SC order parameter has the same overall structure as in the pure TSC state, but with $\Delta_0\simeq 0.174t$ along with $\Delta_{\rm C}\simeq 0.29t$ in the bulk. For both pure TSC and mixed states, $\tilde{h}=0.9552t$ as for the pure TCDW state. The mixed state also hosts two zero modes per surface with $E\simeq 10^{-6}t$. 

\section{\label{sec:conclusions}Conclusions}

In summary, by solving for the superconducting order parameter self-consistently, we have found that the effects of a nonmagnetic impurity in a 2D $s$-wave topological superconductor can vary significantly depending on the material parameters. In particular, the weaker the Rashba SO coupling, or the stronger the Zeeman field, the more $p$-wave-like the system reacts to a nonmagnetic impurity. The midgap excitation bound to the impurity always carries down spin at the impurity site, i.e., spin antiparallel to the direction of the Zeeman field. For relatively strong Zeeman field, however, the spin of the midgap excitation can flip as the impurity potential is varied. We have found two such cases: in one case, the midgap excitation disappears into the gap edge and reappears with opposite spin, and in the other the system undergoes a phase transition of the ground state, whereby the spin-down quasiparticle is bound to the impurity in the ground state and thus the midgap excitation becomes spin up. In this case, a nonmagnetic impurity in a 2D $s$-wave topological superconductor acts exactly like a magnetic impurity (classical spin) in a conventional $s$-wave superconductor. 
The spin flip of the quasiparticle excitation can be understood partially in terms of the TKNN number that corresponds to the shifted chemical potential at the impurity site having the opposite sign to that of the bulk TKNN number. 

We have also shown by solving for each spin component of the Hartree potential as well as the order parameter self-consistently that the Hartree potential effectively reduces the Zeeman field. Furthermore, we have demonstrated in terms of our self-consistent solutions the coexistence of TCDW and TSC in the Abelian phase at half filling. This is analogous to the CDW and uniform superconducting states being the degenerate ground states at half filling in the tight-binding model for conventional $s$-wave superconductivity: the presence of a nonmagnetic impurity, however, lifts this degeneracy in favour of CDW.\cite{Tanaka2000}

In this work, we did not solve for the Hartree potential self-consistently when studying the impurity effects. When the Hartree potential is included, the TSC states presented in Sec.~\ref{sec:impurity} will require weaker pairing interaction and/or SO coupling. 
Further studies of the effects of nonmagnetic impurities and their influence on possible interplay of TCDW and TSC in a 2D $s$-wave topological superconductor will be presented in a future publication.\cite{Goertzen}

\section{\label{sec:acknowledgements}Acknowledgements}

We thank Evan Smith for useful discussions. K.T. is grateful to CCSE at Japan Atomic Energy Agency for hospitality, where part of the research was performed. This work was enabled in part by support provided by WestGrid (www.westgrid.ca) and Compute Canada Calcul Canada (www.computecanada.ca). Some preliminary calculation was also performed on the supercomputing system SGI ICE X at the Japan Atomic Energy Agency. The research was supported by the Natural Sciences and Engineering Research Council of Canada, and partially by JSPS KAKENHI Grant No. 26800197 and the ``Topological Materials Science'' (No. 16H00995) KAKENHI on Innovative Areas from JSPS of Japan.


\end{document}